\documentclass[11pt,twoside]{article}
\usepackage{asp2014}

\aspSuppressVolSlug
\resetcounters

\begin{document}

\title{Using the SourceXtractor++ package for data reduction}

\author{M.\ K\"ummel,$^1$ A.\ \'{A}lvarez-Ayll\'{o}n,$^2$ E.\ Bertin,$^{3,4}$ P. \ Dubath,$^2$ R. Gavazzi,$^{4,5}$ W.\ Hartley,$^2$ and  M.\ Schefer,$^2$ on behalf of the Euclid Consortium}
\affil{$^1$LMU Faculty of Physics, Scheinerstr. 1, 81679 M\"unchen, Germany; \email{mkuemmel@usm.lmu.de}}
\affil{ $^2$Department of Astronomy, University of Geneva, Chemin d'\'Ecogia 16, CH-1290, Versoix, Switzerland}
\affil{$^3$Canada-France-Hawaii Telescope, 65-1238 Mamalahoa Hwy, Kamuela, HI 96743, USA}
\affil{$^4$Sorbonne Universit\'e, CNRS, UMR 7095, Institut d'Astrophysique de Paris, 98 bis bd Arago, F-75014 Paris, France}
\affil{$^5$LAM, Aix-Marseille Universit\'e, CNRS, CNES, Marseille, France}

\paperauthor{Martin~K\"ummel}{mkuemmel@usm.lmu.de}{0000-0003-2791-2117}{LMU Munich}{Faculty of Physics}{Munich}{Bavaria}{81679}{Germany}
\paperauthor{Emmanuel~Bertin}{bertin@iap.fr}{0000-0002-3602-3664}{Sorbonne Universit\'e}{IAP}{Paris}{}{F-75014}{France}
\paperauthor{Marc~Schefer}{Marc.Schefer@unige.ch}{}{Universit\'{e} de Gen\`{e}ve}{}{Versoix}{}{1290}{Switzerland}
\paperauthor{Raphael~Gavazzi}{raphael.gavazzi@lam.fr}{0000-0002-5540-6935}{CNRS - Aix-Marseille Universit\'e}{LAM}{Marseille}{}{13013}{France}
\paperauthor{Alejandro~\'{A}lvarez-Ayll\'{o}n }{alejandro.alvarezayllon@unige.ch}{}{Universit\'{e} de Gen\`{e}ve}{}{Versoix}{}{1290}{Switzerland}
\paperauthor{Pierre~Dubath}{Pierre.Dubath@unige.ch}{}{Universit\'{e} de Gen\`{e}ve}{}{Versoix}{}{1290}{Switzerland}




  
\begin{abstract}
The \textit{Euclid} satellite is an ESA mission scheduled for launch in September 2023. To optimally perform critical stages of the data reduction, such as object detection and morphology determination, a new and modern software package was required. We have developed  \textsc{SourceXtractor++} as open source software for detecting and measuring sources in astronomical images. It is a complete redesign of the original \textsc{SExtractor}, written mainly in C++. The package follows a modular approach and facilitates the analysis of multiple overlapping sources over many images with different pixel grids. \textsc{SourceXtractor++} is already operational in many areas of the \textit{Euclid} processing, and we demonstrate here the capabilities of the current version v0.19 on the basis of a set of typical use cases, which are available for download.
\end{abstract}
\section{Introduction}
\textit{Euclid} \citep{2011arXiv1110.3193L} will cover $15\,000\,{\rm deg^{2}}$ and $40\,{\rm deg^{2}}$ with its Wide Survey and Deep Survey, respectively and is expected to detect and analyse $\approx10^9$ individual objects. To contribute to this task we have developed \textsc{SourceXtractor++} \citep{2020ASPC..527...29K, 2020ASPC..527..461B} as a replacement for and an extension of the original \textsc{SExtractor} \citep{1996A&AS..117..393B} software.
In the \textit{Euclid} processing \textsc{SourceXtractor++} is being used
in the object detection \citep{2015ASPC..495..249K}
and the photometric validation of ground-based data. The software has excelled in the Euclid Morphology Challenge
\citep{2022arXiv220912907E,2022arXiv220912906E}, a recent comparison of several up-to-date model fitting solutions, and \textsc{SourceXtractor++} was selected to perform flexible model fitting with S\'{e}rsic and Disk+S\'{e}rsic models to all \textit{Euclid} objects.
\textsc{SourceXtractor++} is distributed\footnote{\url{https://github.com/astrorama/SourceXtractorPlusPlus}} as an independent package via source code or various binary distributions (conda, Fedora/Centos RPMs) to the astronomy community.
\section{Test Data Set}
\textsc{SourceXtractor++} can be used in a variety of ways to work on different kinds of input data (single band, multi-band, data cube). To demonstrate this flexibility we have assembled a series of use cases \footnote{\url{https://cloud.physik.lmu.de/index.php/s/3K4KemBsw5y9yqd}}. The use cases contain all necessary input data, the configuration files and, as a reference, the expected results. The exercises are organized in several sub-directories:
\begin{itemize}
\item {\tt 01\_basics}: Basic operations such as generating a blank ASCII configuration file, listing all available object properties or performing a simple object detection as in \textsc{SExtractor2} are shown. Different image extensions for detection and pixel weights are accessed. S\'{e}rsic models are fit to all detected objects. All settings for the model fitting are done in a python configuration file which is executed at the start of the fitting process.
\item {\tt 02\_mband}: Disk+Bulge models
are simultaneously fit to three measurement images while using a fourth image for the object detection. The measurement images are grouped according to the filter information provided in the header, which results in one independent flux measurement per group. The images coming from JWST cover different areas and have diverse pixel scales. All relevant positions and parameters are negotiated with the image WCS. Arbitrarily defined dependent parameters can be pushed to the output. The usage of python for the configuration file facilitates pre-processing the data, e.g. band selection or keyword adjustments;
\item {\tt 03\_priors}: This example was taken from \citet{2022A&A...666A.170Q} and illustrates the refined multi-band fitting of a bright galaxy. With priors the fit results can be constrained to avoid the influence of neighbour objects such as bright stars. Priors in connection with dependent parameters allow large flexibility such as fitting band specific positions while maintaining meaningful results by keeping the position differences small.
\articlefigure[width=1.\textwidth]{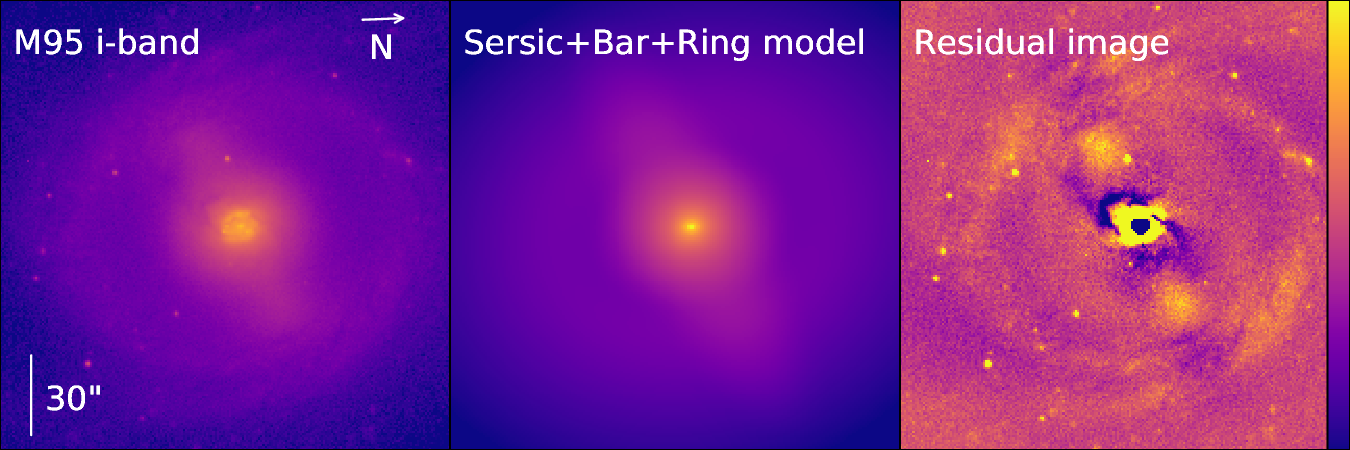}{fig1}{Fitting a {\tt S\'{e}rsic+Bar+Ring} model to M95 with user defined functions. The SDSS i-band image is shown in the left, the fitted model in the middle and the residual image in the right sub-panel.}
\item {\tt 04\_user\_defined}: The functional forms, S\'{e}rsic, exponential disk,  deVaucouleur Bulge and point-like are provided in \textsc{SourceXtractor++};
the exercise shows how users can define an ONNX model to define customized functional forms. \textsc{SourceXtractor++} uses the ONNX Run time\footnote{\url{https://onnxruntime.ai}} to evaluate the ONNX models and fit higher level features such as rings or bars. Figure \ref{fig1} shows the fit of a S\'{e}rsic+Bar+Ring to the SDSS i-band data of M95.
\item {\tt 05\_cubes}: \textsc{SourceXtractor++} can run on multi-extension FITS images and even on data cubes. The example uses a broad-band image for detection and then performs model fitting on several image groups selected from layers in a data cube from the MUSE/ESO instrument observing the Abell 2744 galaxy cluster\footnote{\url{http://muse-vlt.eu/science/a2744}}. The layers are selected by wavelength range in the python configuration file using the WCS of the cube.
\item {\tt 06\_extended\_assocs}: \textsc{SourceXtractor++} has implemented an extended association mode which allows the user to funnel external object properties such as starting positions or shape parameters into the model fitting process.  This mode is a first step towards disentangling the detection and the measurement stages. Once fully developed it will be possible to run the measurements solely based on objects properties (positions, segmentation) provided to \textsc{SourceXtractor++}.
\item {\tt 07\_python\_control} This exercise runs currently only in a special branch of the v0.20 development version. The sequence of processing steps (detection, grouping, measurement) is executed directly from Python. Throughout the processing the user has access to the objects and other processing items and can apply selections or run their own analysis code on the objects. \textsc{SourceXtractor++} can then integrate completely into the more complex processing pipeline of the user.
\end{itemize}
\articlefigure[width=.8\textwidth]{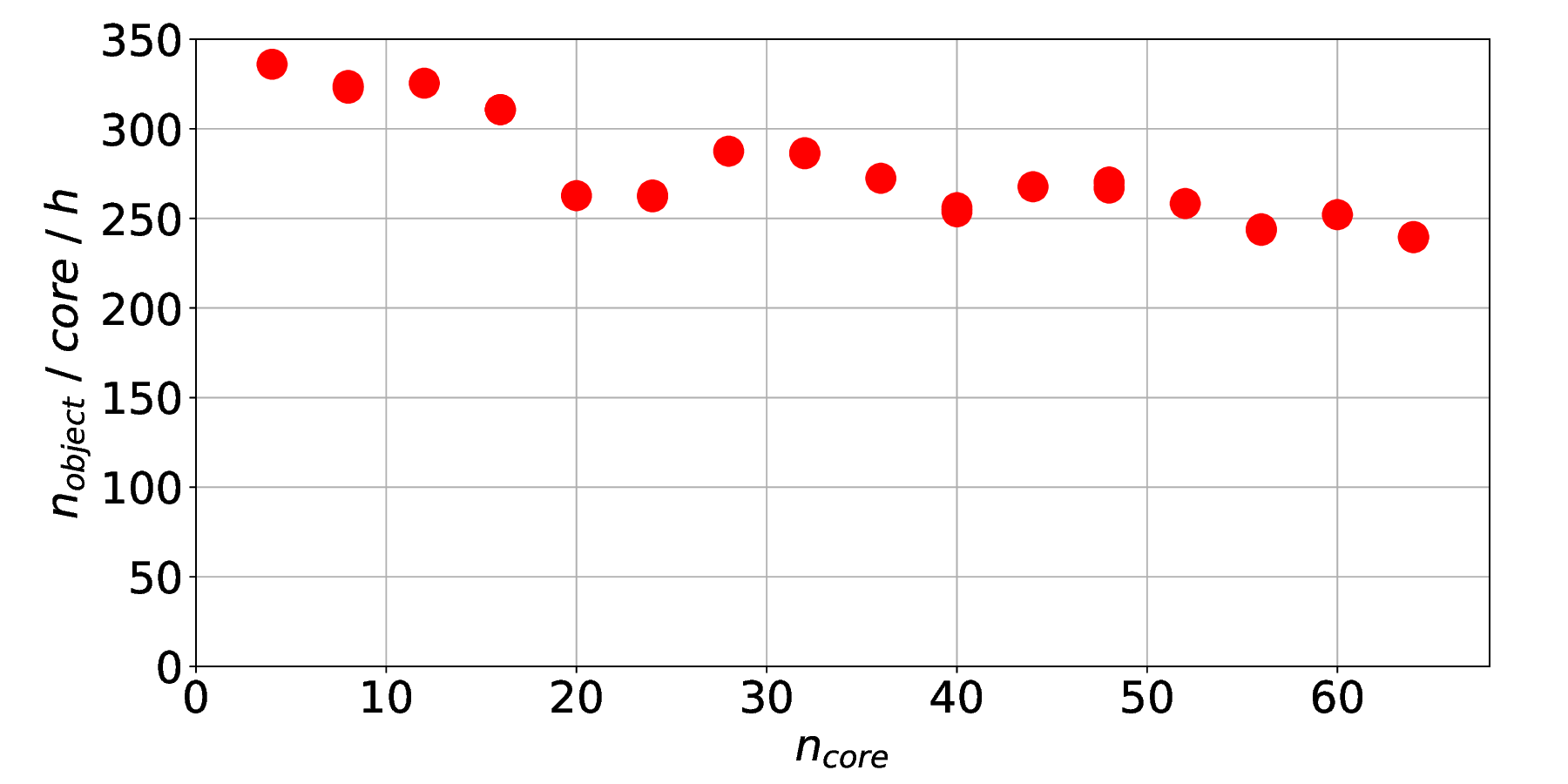}{fig2}{The efficiency of the multi-core processing in \textsc{SourceXtractor++}.}
\section{Status and Outlook}
\textsc{SourceXtractor++} has been available for a few years and has also picked up users outside of the \textit{Euclid} community. In addition to the \textit{Euclid} requirements we designed \textsc{SourceXtractor++} as a general processing tool to offer novel possibilities for reducing data. The efficient parallel processing in \textsc{SourceXtractor++} is evidenced by the rather flat curve in Figure \ref{fig2}, which shows the number of objects processed per hour and core as a function of the number of processors assigned to the task. \textsc{SourceXtractor++} contains many of the features we had in mind when starting the project. Future developments will concentrate on becoming more \textit{Pythonic} and separating detection and measuring, as indicated in the last two exercises. We are certain that \textsc{SourceXtractor++} offers new possibilities for reducing data both within and outside of the \textit{Euclid} project.
\acknowledgments The Euclid Consortium acknowledges the European
  Space Agency and a number of agencies and
  institutes that have supported the development of {\it Euclid}, in
  particular the Academy of Finland, the Agenzia Spaziale
  Italiana, the Belgian Science Policy, the Canadian Euclid
  Consortium, the French Centre National d'Etudes Spatiales, the
  Deutsches Zentrum f\"ur Luft- und Raumfahrt, the Danish Space
  Research Institute, the Funda\c{c}\~{a}o para a Ci\^{e}ncia e a
  Tecnologia, the Ministerio de Ciencia e
  Innovaci\'{o}n, the National Aeronautics and Space
  Administration, the National Astronomical Observatory of Japan,
  the Netherlandse Onderzoekschool Voor Astronomie, the Norwegian
  Space Agency, the Romanian Space Agency, the State Secretariat
  for Education, Research and Innovation (SERI) at the Swiss
  Space Office (SSO), and the United Kingdom Space Agency. A
  complete and detailed list is available on the {\it Euclid} web site
  (\texttt{http://www.euclid-ec.org}).
\bibliography{F06}
\end{document}